# Collaboration Framework in the EViE-m Platform

Kostas Kapetanakis[1], Haroula Andrioti[2], Helen Vonorta[3], Marios Zotos[4], Nikolaos Tsigkos[5], Ioannis Pachoulakis[6]

*Department of Applied Informatics and Multimedia, Technological Educational Institute of Crete*
*Heraklion, Crete, Greece*
[1]kapekost@epp.teicrete.gr, [2]mtp1@edu.teicrete.gr, [3]tp3289@edu.teicrete.gr,
[4]epp2355@edu.teicrete.gr, [5]epp2211@edu.teicrete.gr, [6]ip@epp.teicrete.gr

*Abstract*— **Within the context of a 3D interactive strategy game, the EViE platform allows participants to unlock game features using their knowledge and skills in various thematic areas such as physics, mathematics, etc. By answering questions organized by Educational Objective in stratified levels of difficulty, users gather points which grant them access to desired world elements. Richer world components become increasingly more difficult to access, so that a players' individual (or cumulative if in a group) knowledge, ability and / or dexterity is directly reflected by the level of complication of their virtual world. In the present article we report on the communication architecture of the platform and focus on framework components that allow group activities such as cooperation (within the group to facilitate e.g., collaboration on more difficult problems), (inter-group) competition as well as practice and skill honing activities (in single or in multi-player mode).**

## I. INTRODUCTION

Modern technological advances, especially in the area of virtual reality (VR), offer unique real world emulation where people directly visualize the consequences of their actions. Virtual environments where users can build up their own virtual worlds assist in extending knowledge and spatial memory [1] and are more effective in practicing material [e.g., 2]. However, a shift of focus from simple task completion to building conceptual actions may be necessary to change the perception of knowledge [3]. In addition, adaptive systems that are based on students' learning ability and learning style improve their learning achievements. In fact, adaptation according to learning style accelerates the learning process [4]. Other attempts, such as in [5] extend traditional e-Learning systems by providing interacting platforms for students to contribute and socialize. By adopting an agent-based swarm intelligence system to manage the group learning path and resources, [6] show that various characteristics of system behaviour correlate with user satisfaction and with learning benefits, supporting that users of such learning environments outperform users of more generic web-based learning environments. Educational games combine mental ability, fast thinking and effort to make the right choices under pressure in order to implement strategies to overcome obstacles in the game [7].

The EViE platform is being developed by a small community of programmers and researchers aiming to provide a robust virtual educational environment to aid the learning process [8]. The platform in particular [9-11], implements a strategy game that uses mathematical knowledge to unlock the more advanced features and discriminate among users based on mathematical ability and dexterity [12]. The platform offers a virtual environment (a sample runtime screenshot appears in Fig. 1) based on 3D graphics and can be adapted in various educational areas. We have already adopted EViE for high school level Mathematics and refer to it as EViE-m. The core of the platform is written in Java and utilizes OpenGL in conjunction with Xj3D libraries to display the 3D world and support user interactions.

In addition, a Network Module (NM) is implemented in the main platform to provide multiplayer gaming experience via peer-to-peer connection between network-enabled instances. An initial implementation using JXTA peer-to-peer connection [13] surfaced numerous JXTA-related bugs and was abandoned in favour of a custom-made peer-to-peer connection framework, which is described in this paper. Platform users are called to construct a virtual city using a variety of structures such as houses, hospitals, police stations, shops and cinemas, access to which is granted through answering correctly multiple-choice questions or work out short problems. Some of the structures have prerequisites, meaning that access to a desired construct is granted through collecting points and adding lower-level constructs. Questions are organized by difficulty level and also by Educational Goal.

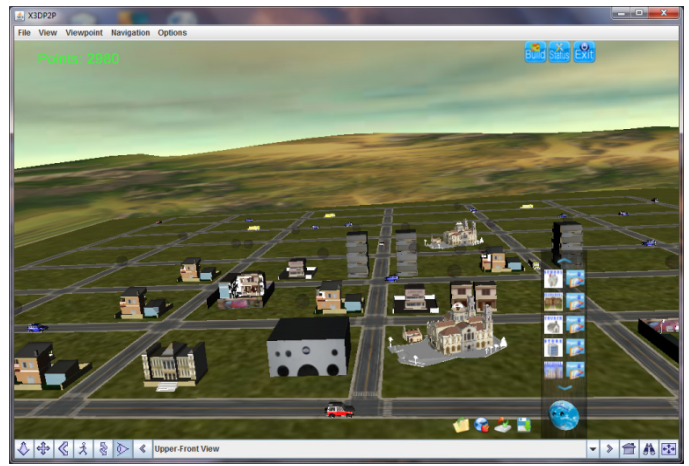

Fig. 1  Sample user view of the EViE-m environment

The Xj3D browser is an open source, cross platform X3D player [14]. Central rendering algorithms operate within Xj3D libraries which utilize OpenGL. Scene navigation operations and the triggering events from external resources such as mouse and keyboard are managed by Xj3D components. Game start-up creates the virtual environment and initializes the proper components to utilize user interactions. Users may select single-player or multi-player mode through the main panel. In single-player mode, a user starts with zero points and uses main panel buttons to pick a desired construct to insert to the world, after which event a question and possible answers appear. A correct answer allows instantiating the structure in the world. The more questions are answered correctly, the more points are accumulated and more structures are unlocked. As the game progresses more advanced questions appear.

In multi-player mode, the same logic is implemented for a world which is now common to the entire group. Additionally, a Text Communication Module provides inter-communication between users to facilitate the collaboration. In such a context, individual user and team scores are maintained.

The paper is structured as follows: Section II describes the main platform modules and their functionalities and Section III discusses platform setup. Section IV provides a development guide and, finally, Section V discusses core classes and the game logic. We conclude with a presentation of our future plans.

## II. MAIN MODULES

The EViE platform is built around modules which are combined during initialization and game play. For instance, the Question Manager (QM) module provides proper XML files for the Educational Objectives and the Questions per such objective. Entering multi-player mode enables the Team Selection (TS) module which controls peer creation and team joining or leaving, the Intra-Communication (IC) module that processes the intra-system messages in order to provide server-independent peers, and the Text Communication (TC) module which combined with the Information (I) module control an Instant Messenger and a floating window that contains two tabs: the instant messenger and the information panel.

### A. The Question Manager module

As referred in [16], the EViE-m platform implements a stand-alone application written in java to manage the available questions. The tutor can add/edit/delete an Educational Objective (EO) and the questions under each EO. This module can be initialized externally, independently of the main platform and edits the proper xml files corresponding to the input data for the platform. A subfolder named "questions" encloses the questions.xml and the lessons.xml files which hold the data for the available questions that appear to the students during the game-play. The interface to each question leads the question or problem statement to appear on top of a list of selections where the later are each assigned to a radio button in a single radio button group. On the right side of the panel an optional accompanying image appears.

The main class for the QM module is QuestionEditor and is included in the questionManager package. JFrames are extended in this class to create a user-friendly interface to help the tutor manage the available questions. The application provides all necessary fields to create the required XML files with the Questions and the Educational Objectives for the EViE-m platform.

### B. The Team-Selection Module

According to [6], the Team selection panel appears when the user selects the network connection button located at the bottom of the main panel (see Fig 2). The system broadcasts a message to inform all the available, network-enabled, clients about the new online-user. The peers receiving this message reply with a list of the available teams. Each peer works as a host and as a client at the same time, thus the platform is server-independent. Each received message triggers events even at the sender's instance. The network panel is triggered once a message requests the available teams which have arrived back to the sender (loopback message). Furthermore, the application is aware of the network availability so that, if the message never returns, then the network panel is not triggered.

The network selection panel provides the user with a list of the available teams (once each peer has announced its list of teams). Additionally, an option to enter the name in a text field is provided. When the student types a team name which is not in the list, that particular team is created when the "join" button is pressed.

### C. Intra-communication Module

When the network module is enabled, users can cooperate from different terminals to construct the city following the same procedure. The "network" button on the bottom left side of the panel triggers the IC module. A "Welcome" message is broadcasted from the peer with a notation of group/team request. This message, as all messages, arrives back to the sender and to each network-enabled peer. When a peer receives its own "Welcome" message with group request notation, the Team Selection panel is triggered. At the same time the content of this panel (the list of groups) is generated by the response messages arrived by all other peers containing a list of the groups each peer knows about. At this point the new peer which broadcasted the "Welcome" message generates a list of the available groups and the user is ready to join a team or generate a new team. When the user requests to join a team, a new type of "Welcome" message is broadcasted with a notation of "Group selection" that contains the name of the team. Then, a synchronization process takes place.

During synchronization, each peer informs the newly connected peer about the constructed structures. During this process a new list is generated in the system of the newly connected peer, containing the IP address of each peer in the same group. Furthermore, when the process goes to the next step, the other peers broadcast messages for the new peer to invoke the necessary processes and construct the same buildings at the corresponding rotation and translation values (placement axis, and rotation degrees).

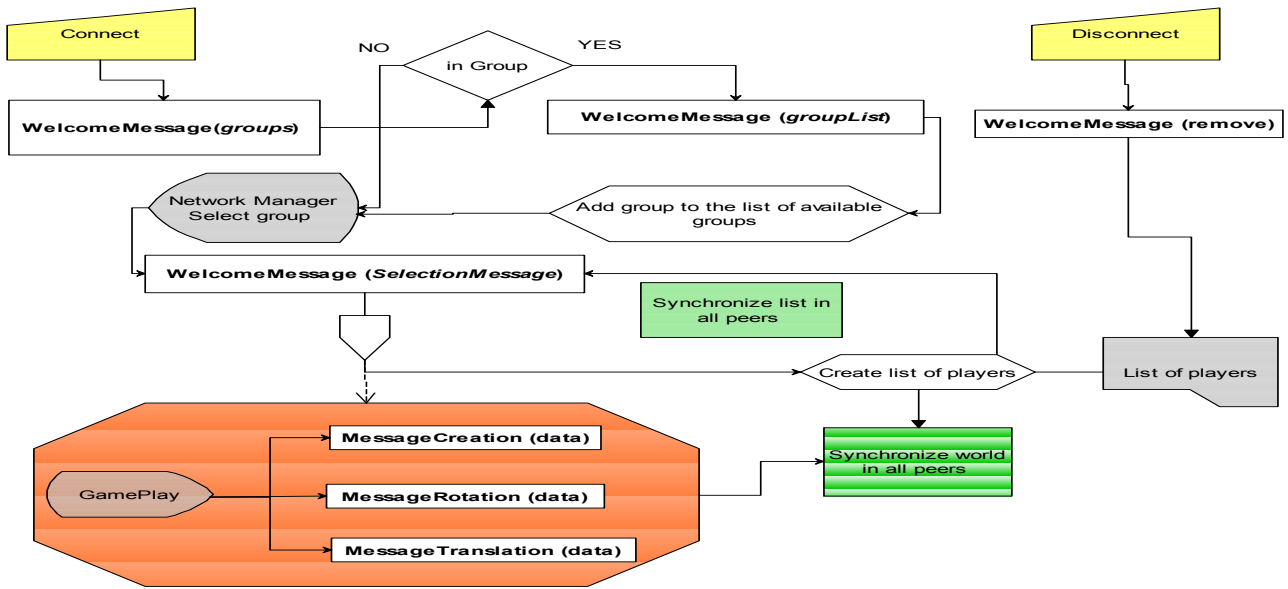

Fig. 2  Team selection panel

Another type of "Welcome" message is also used to notify peers about a disconnection. When a user triggers the "exit" button or the "disconnect" button, a message is broadcasted with a notation of removal. Then each peer removes the IP address from the list of group members. As a user interacts with the game, system messages are generated to update world status on all other peer instances. Those messages are tagged as "System Messages". If the sender belongs to the same group, then the message is processed, otherwise it is discarded. Since a message may arrive more than once, depending on the number of active peers, the platform maintains an array to identify each structure based on a unique ID message and disposes messages that would trigger a rebuild of the same structure.

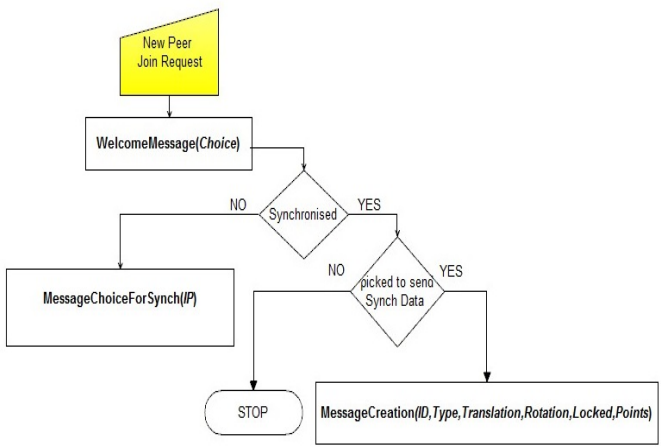

Fig. 3  Peer/World synchronization process

Synchronization is activated only when a new peer enters a group which already has more than two members. As shown in Fig.3, a flag is used in order to create a filter showing if the peer actually needs to be synchronized with the world. In that case the message "choice" is sent which contains the IP address of the peer picked from the current group (team). Only the peer that is addressed broadcasts messages for the new peer to construct the same virtual world.

### D. Text Communication Module and Information Module

A floating window appears at the bottom left side of the screen when the network mode is enabled (Fig. 4). A short message appears on the floating window, notifying the user to click. When the clicks this area, the window rolls out to display the "status" and "chat" tabs. The "status" tab holds information about the multiplayer mode, the points that a particular user has and the contribution points (the points that this user has contributed to the team). Furthermore, the second panel, "chat", provides an instant messenger that delivers messages to other team members. The student can set and change the username that will appear before the instant messages are delivered from the current terminal. Each username is prefixed by the last digits of the IP address of the peer. While the users exchange messages, a log file named "chat.xml" is created in a subfolder named "models".

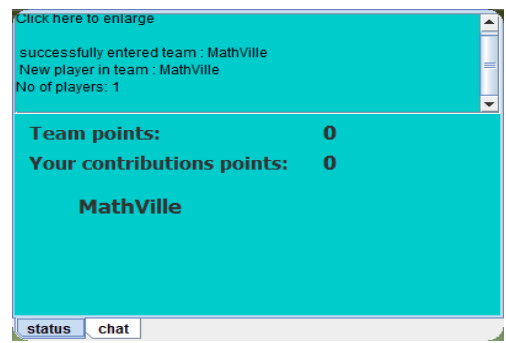

Fig. 4  The "Text Communication" and "Information" modules

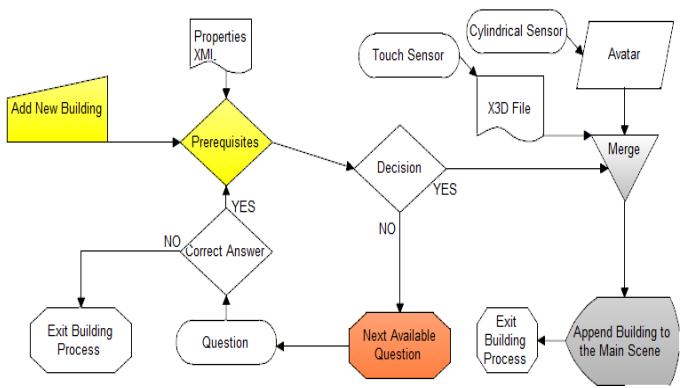

Fig. 5 Construction Module

*E. Construction Module*

Using the main panel, users can request a construct to be built along the lines of the process shown in Fig. 5. Each construct has a "properties" button that displays the points it offers in each area, and the prerequisites. The platform validates the prerequisites of the requested construct to trigger the Question panel. In this panel, the user has to pick the correct answer. When the answer is submitted, the platform verifies the answer. If the answer is wrong, then the user can request a new one. If it is correct, the process goes on to build the construct. The construct is an external X3D file which is transformed during the merging procedure. In the requested X3D model (the structure) a touch sensor and a cylindrical sensor are configured to enable the user interaction with this construct. The touch sensor is placed on top of the 3D model, so that the user may relocate the construct. The cylindrical sensor is embedded as a dynamically generated additional 3D object (avatar). The avatar absorbs the drag gesture of the mouse, and provides the rotation values to its parent 3D model (the structure). The composite object representing the structure with its avatar is placed in the world. The user may then move and rotate the added structure using the mouse.

The X3D files of the structures are saved in a "models" folder, each in a appropriately named subfolder along with its textures and any additional information. Each structure follows the same process to adjust the external X3D file and insert it in the main world. The final object is assigned a unique ID in order to avoid rebuilding it.

## III. INITIALIZATION PROCESS

The platform is provided with a .bat file named "run" that creates the proper environment in order to run the game outside of the development environment (Fig. 6). If we want to run the platform in windows we use this file to open the command and initialize the Xj3D browser which is a .jar file. On a UNIX platform we can normally run the jar file with a java command. The initialization takes place in the Xj3Dbrowser class which creates a JFrame in full screen mode. In the main panel of this frame, an X3D world is rendered which is stored in the background.x3d file. X3D, on the other hand, can implement ECMAscript and we used this feature to invoke our main platform. The ECMAscript can run both java and JavaScript. In our implementation, an external source is defined for this script, which is a compiled java file (...url="'../build/classes/BuildingPanelSC.class'">).

The virtual world has now been initialized with the proper sensors and the corresponding 3D environment. Additionally, in this background.x3d file, the main panel of the platform is stored and attached to the current view point in order to appear and accept the user's input. All ROUTES and fields, for and from the script are initialized in the current script node. When the main class of the platform is initialized, each field and route from the java core is bound to the corresponding field and route of the virtual world. Therefore, data from the virtual world may pass to the java core for further process and the output of each algorithm can be routed to the virtual world. A harmonic operation takes place between both technologies to create a hybrid virtual environment based on X3D and Java objects. Finally, in the folder named models, an XML file is stored to hold the data during the game-play. An XML file named "w_status.xml" is maintained during game-play to hold the current game status, the collected points and the constructed buildings. The same process is implemented to calculate the current level depending on data from external xml files.

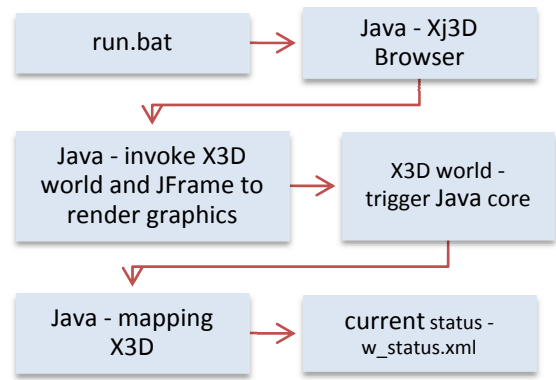

Fig. 6 Initialization process

The platform is implemented in JAVA in cooperation with Xj3D player. The EViE platform code and data files are distributed as a Netbeans project. A container folder named "HouseGame" at the root of C:\ includes the resources and additional .jar and .dll libraries. The libraries include both xj3d and OpenGL in the version the game was developed. The main class in order to run the platform is in the Xj3DBrowser.java file.

## IV. CORE CLASSES

The core of the platform is developed in java. The Xj3DBrowser class (Fig. 7) is responsible for starting the game. The JFrame has been extended to build the main frame of the platform which hosts the navigation controls as well as the menu bar for common actions such as capturing a single frame or sequence of frames. The central area of the frame hosts the main 3D world to be rendered, which is saved as background.x3d file.

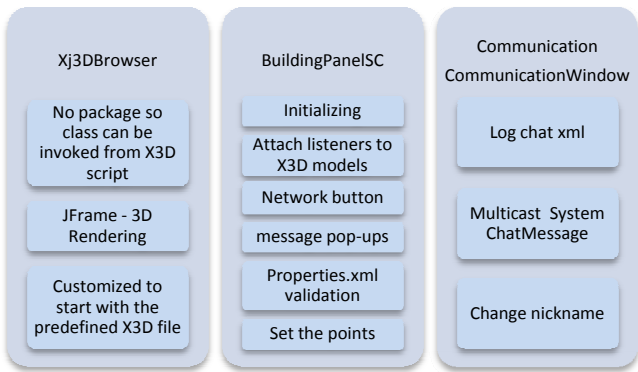

Fig. 7 The "Core" classes

The initiated X3D world then triggers the core of the platform under the BuildingPanelSC class. At first, a process to bind the X3D world fields with the corresponding java fields, takes place. In addition, this class implements the main menu buttons actions. Each listener is initialized and invokes the rest of the modules corresponding to the user's actions.

The CommunicationWindow class is responsible for the small panel that appears in multiplayer mode. On the bottom left side of the screen, this class generates a floating JFrame which after being clicked displays the current player contribution and team total points as well as a second tab for team communication using the chat module. The class also creates a log file with the exchanged text messages between the players, as well as a text area to nickname players.

The Networklayer package (Fig. 8) includes the MulticastNet, a background class for the network layer module responsible for network connectivity and managing incoming and outgoing messages. This class provides LAN multiplayer support and is being enhanced with a module for internet-based peer communication.

The WorldManager core class manages messages that are locally generated or originate on other peers and updates world status appropriately. In addition, every message is further processed to provide the corresponding world construct or interaction of the user. Messages that contain information about the remote and local constructs are separated according to functionality, e.g., if they refer to an already built construct, translation/rotation values are passed. Furthermore, when a construct does not belong to the current peer, a lock method is invoked to neutralize translation/rotation sensors in order to avoid manipulation from users other than its creator.

When the player requests a construct to be built, the corresponding messages trigger several events (Fig. 9), which are processed depending on the game's current status. When the message arrives at the Building class, the system merges the x3d file of the construct (including the texture files) and calculates the new status depending on the properties file of the construct. In addition, the Building class appends a series of listeners and Route objects to the world to enable the construct movement and rotation. This process implements functionalities from other related classes such as the BuildingNurse. Following completion of this the construct appears in the virtual world with a small avatar to handle rotations. World constructs can be moved using a plane-sensor and corresponding Route nodes which transfer the mouse-dragging values to the parent node, to realize construct relocation. On the other hand, when the mouse is dragged over the avatar object, the construct is rotated. An avatar object is dynamically appended on each built construct, and a cylindrical sensor (X3D API) is attached onto it. The sensor transfers the mouse-drag values to the parent construct via a dynamically added Route node. During construct build up, the Housemanager class reads the required data and proceeds only following proper validation.

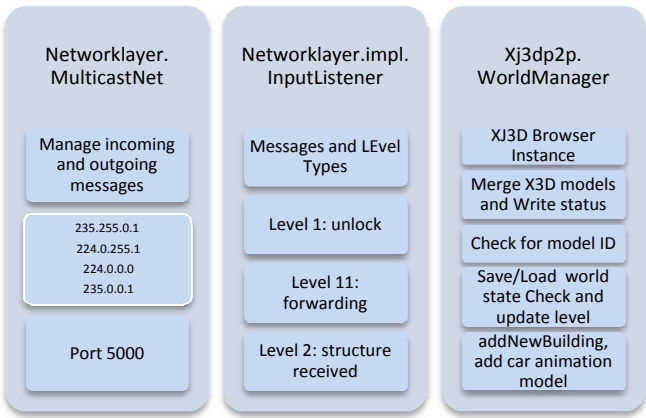

Fig. 8 Network Layer and core Manager class

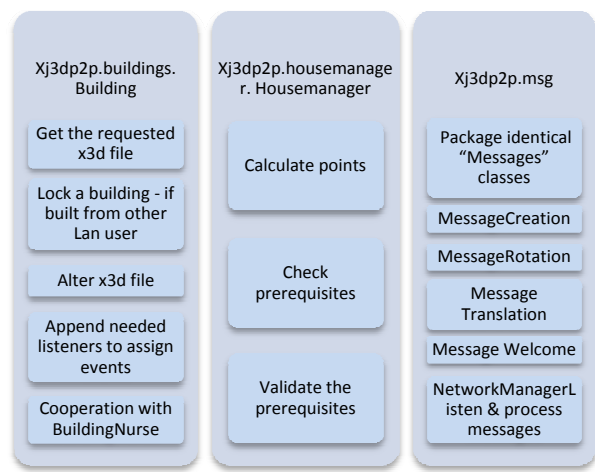

Fig. 9 Building Procedure and Messages

The inputListener class on the other hand, manages the incoming messages in a queue and forwards them to the proper subsystem of the platform (chat module, network setting module, etc).

The msg class acts as an interface to message exchange between modules. The platform filters newly created messages as well as the ones intended for redistribution. The main message types are: a) rotation, b) translation, c) creation and d) handshake. The rotation message holds a pointer to the intended construct, the rotation axis and the rotation angle. A translation message also specifies the construct to be moved as well as its final position. A creation message is a combination of a rotation and translation message. Finally, handshake messages are tagged "Message Welcome". The msg class implements the algorithm that redistributes the messages as a NetworkManager subclass. This network manager filters the messages that are broadcasted from other peers and points to further processing.

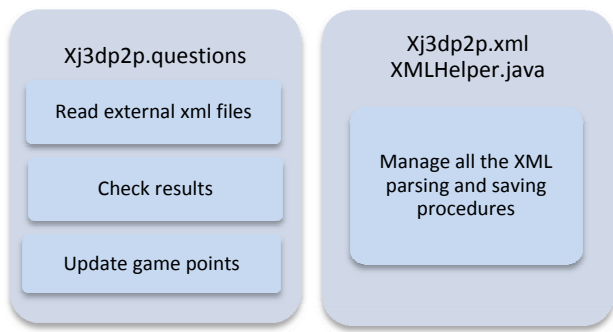

Fig. 10 Question Procedure and XML Helper

The classes of the question module are implemented to listen to requests and randomly select the next question (Fig. 10). In addition, this class checks the answer provided by the user in order to continue with the merging process and display the requested construct into the main world and the calculated contribution points. The question appears in front of the user during game-play in the form of a simple JFrame containing data based on the corresponding XML questions file. The questions are selected randomly from question set of the current difficulty level.

The platform requests to read and write XML text files during the game play. XML is used to prescribe X3D files, save the world status, prescribe the properties of each construct and organize questions under each Educational Objective. The XMLHelper class is invoked whenever the platform needs to write or read data to/from an XML file. The XML documents in JAVA are constructed using the w3c.dom.Document class.

## V. DISCUSSION AND FUTURE PLANS

The EViE platform provides a cross-platform, user-friendly 3D interactive virtual learning environment where students can create a virtual city independently or in groups, be rewarded with special constructs as they progress to more difficult material. At the time of writing of this article we have run tests with small student groups to locate bugs and receive user feedback. Early next school year we intend to run larger-scale tests. Word in progress includes providing a web-based clone of the platform based on HTML5, Javascript and a Java websockets server to enable running using any browser and without additional software installation.

In addition, specific modules are being planned to provide tutors with a complete management system and observation techniques to monitor the over-all process of the game for each group. The adapted technologies for the web-based version of the platform and the tutoring module should provide an interoperable system where smart-phone based monitoring modules should be able to plug in.